\documentclass{article}

\usepackage{PRIMEarxiv}

\usepackage[utf8]{inputenc} 
\usepackage[T1]{fontenc}    
\usepackage{hyperref}       
\usepackage[numbers]{natbib}
\usepackage{siunitx}
\usepackage{url}            
\usepackage{booktabs}       
\usepackage{amsfonts}       
\usepackage{nicefrac}       
\usepackage{microtype}      
\usepackage{lipsum}
\usepackage{fancyhdr}       
\usepackage{graphicx}       
\graphicspath{{media/}}     

\title{{Stabilizing Laplacian Inversion in Fokker--Planck Image Retrieval using the Transport-of-Intensity Equation}
}

\author{
  Samantha J Alloo and Kaye S Morgan \\
  School of Physics and Astronomy, \\
  Monash University, \\
  Victoria, Australia\\
  \texttt{samantha.alloo@monash.edu} \\
}

\begin{document}
\maketitle

\begin{abstract}
X-ray attenuation, phase, and dark-field (multimodal) images provide complementary information to one another. Different experimental techniques can capture these contrast mechanisms, and the corresponding images can be retrieved using various theoretical algorithms. Our previous works developed the `Multimodal Intrinsic Speckle-Tracking' (MIST) algorithm, which is suitable for multimodal image retrieval from intensity data acquired using a speckle-based X-ray imaging (SBXI) technique. MIST is based on the X-ray Fokker--Planck equation, requiring the inversion of derivative operators that are often numerically unstable. These numerical instabilities can be addressed by employing appropriate regularization techniques, such as Tikhonov regularization. However, the output from such regularization is highly sensitive to the choice of the Tikhonov regularization parameter, making it crucial to select this value carefully and optimally. In this work, we present an automated iterative algorithm to optimize the regularization of the inverse Laplacian operator in our most recently published and most general MIST variant, addressing the operator's instability near the Fourier-space origin. Our algorithm leverages the inherent stability of the phase solution obtained from the transport-of-intensity equation for SBXI, using it as a reliable ground truth for the more complex X-ray Fokker--Planck-based algorithms that incorporate the dark-field signal. We applied the algorithm to an experimental SBXI dataset collected using synchrotron light of a sample comprised of four different rods. The phase and dark-field images of the four-rod sample were optimally retrieved using our developed algorithm, eliminating the tedious and subjective task of selecting a suitable Tikhonov regularization parameter. The developed regularization-optimization algorithm makes MIST a more user-friendly multimodal-retrieval algorithm by eliminating the need for manual parameter selection. We anticipate that our optimization algorithm can also be applied to other image retrieval approaches derived from the Fokker--Planck equation, such as those designed for free-space propagation techniques, as they also involve the unstable inverse Laplacian operator.
\end{abstract}

\vspace{10pt}
\textbf{\textit{20.12.2024}}

\vspace{6pt}

\section{Introduction}
Attenuation, phase, and dark-field are three distinct contrast channels in X-ray imaging, each providing unique insights into a sample's properties. The attenuation image reveals structural density by quantifying how many X-ray photons are absorbed or scattered out of the primary beam. In contrast, the phase channel captures X-ray refraction within the sample, allowing materials with weak attenuation (i.e., those that are almost invisible in attenuation images) to be distinguished \cite{endrizzi2018x}. The dark-field channel visualizes structures that induce phase variations smaller than the system's spatial resolution, thus offering information about features that are otherwise undetectable in attenuation or phase images. Numerous studies highlight the broad applicability of these three contrast channels across different fields \cite{miller2013phase,ludwig2018non,gassert2021x,shimao2021x,blykers2022exploration,he2024nondestructive,schaff2024feasibility,glinz2024comparison}.
\\
Several multimodal experimental techniques have been developed to recover attenuation, phase, and dark-field images, each offering specific advantages and limitations. For example, using free-space propagation \cite{gureyev2009refracting,leatham2023x}, grating-interferometry \cite{pfeiffer2008hard}, edge-illumination \cite{Endrizzi2014}, two-dimensional (2D) single grid \cite{Wen2010,bennett2010grating}, and speckle-based X-ray imaging (SBXI) \cite{zdora2018state}. Each technique employs a unique image acquisition procedure, exploiting slightly different physical changes in the imaging system to capture multimodal sample information. Then, an appropriate retrieval algorithm is required to extract multimodal signals from the experimental intensity data. These algorithms decode the contrast seen in raw images and how that contrast relates to the sample's attenuation, phase, and dark-field characteristics based on the specific setup.
\\
\begin{figure}[tb]
    \centering
    \includegraphics[width=0.75\linewidth]{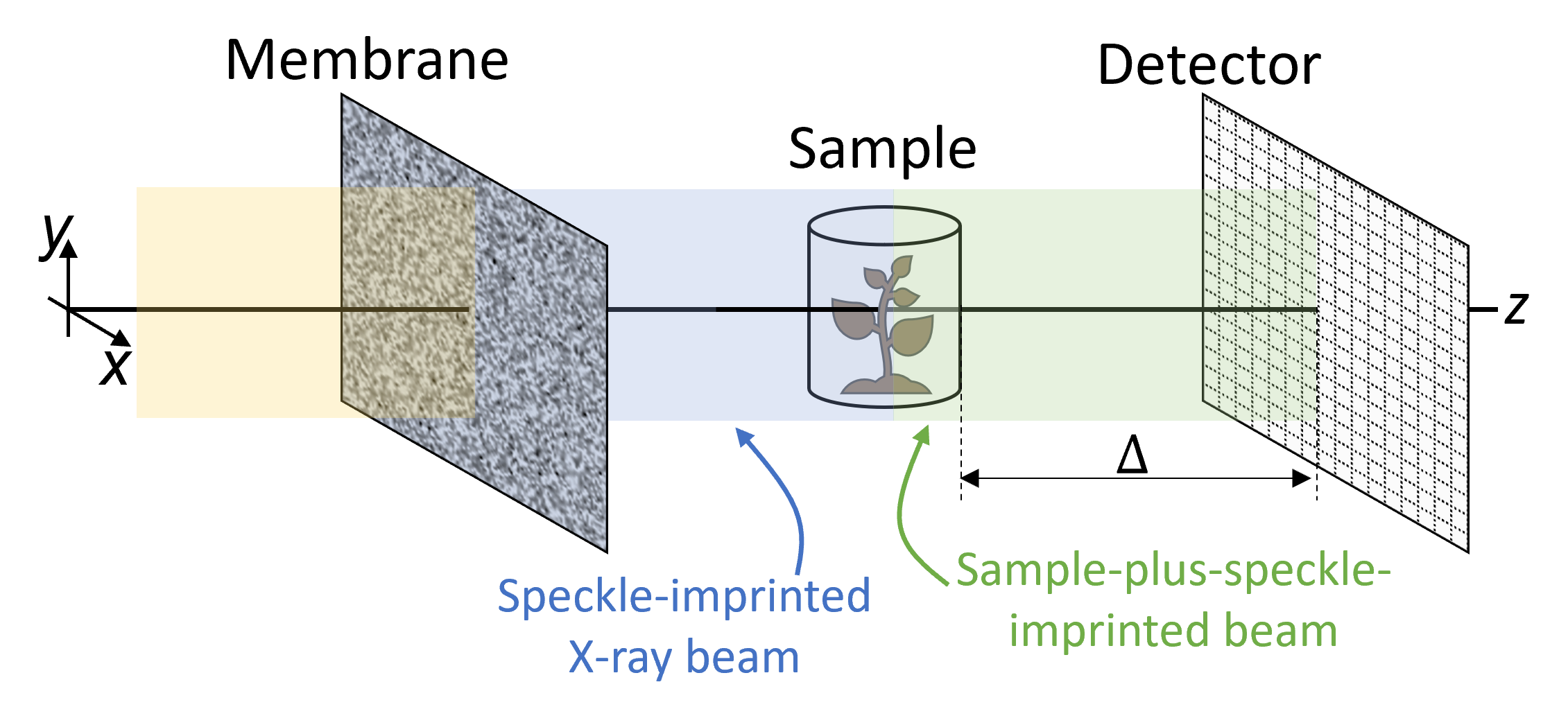}
    \caption{Speckle-based X-ray imaging (SBXI) setup with a paraxial monochromatic X-ray wavefield. A spatially random membrane imprints speckles into the X-ray wavefield to generate a reference-speckle pattern $I_R(\textit{\textbf{r}})$ at the detector. When a sample is introduced, a sample-plus-speckle image $I_S(\textit{\textbf{r}})$ is formed. All images are captured at the detector, positioned at a distance $z = \Delta$ downstream from the sample.}
    \label{fig:setup}
\end{figure}
Our research group has been developing the so-called Fokker--Planck equation (FPE) for paraxial X-ray imaging \cite{paganin2019x,morgan2019applying}, which can also be thought of as a diffusive extension to the transport-of-intensity equation (TIE) \cite{teague1983deterministic}. In this context, the X-ray imaging FPE models how the intensity of an X-ray wavefield, refracted and diffused by a sample, evolves as it propagates along the optical axis. The TIE can be viewed as a simplified version of the Fokker–Planck equation, neglecting X-ray diffusion caused by the sample. Specifically, the  TIE is a mathematical framework that describes how an X-ray wavefield's intensity changes with propagation due to a phase shift introduced by some refracting object. The equation has proven successful in solving the phase-retrieval problem (recovering just phase information), being the basis of the widely adopted Paganin method \cite{paganin2002simultaneous}, which is suitable for free-space propagation phase-contrast X-ray imaging data, and has been implemented in both light microscopy \cite{kou2010transport,waller2010transport,gupta2020single} and electron \cite{streibl1984phase,bajt2000quantitative,petersen2007tem,liu2011projected} microscopy. To also include sample-induced X-ray diffusion, described by an effective diffusion coefficient, the X-ray FPE \cite{paganin2019x,morgan2019applying} should be used instead of the TIE to model the forward problem. The X-ray FPE has been successfully applied to both free-space propagation \cite{leatham2023x, leatham2024x, ahlers2024x} and SBXI \cite{pavlov2020x,alloo2021speckle,pavlov2021directional,alloo2022dark,beltran2023,alloo2023m, magnin2023dark}, modeling the forward problem and enabling the retrieval of multimodal images from experimental data by solving the corresponding inverse problem. In terms of SBXI, which is the focus of this paper, the FPE establishes a relationship between the reference-speckle image and sample-plus-speckle image. The reference-speckle image is captured when a spatially random membrane is placed in the beam path, while the sample-plus-speckle image is recorded when 
a sample is introduced into this reference-speckle field. The experimental set-up of SBXI using a monochromatic paraxial X-ray source, which can be achieved at synchrotron facilities, is shown in Fig.~\ref{fig:setup}. We have been developing an FPE-based multimodal retrieval algorithm for SBXI, known as Multimodal Intrinsic Speckle-Tracking (MIST) \cite{pavlov2020x}. Since its inception in 2020, several MIST variants have been published, each aiming to optimize the solution to the multimodal inverse problem. The ultimate goal is to solve the SBXI FPE fully--or with minimal assumptions about sample properties or experimental conditions--in the most mathematically stable manner, thereby enabling the recovery of high-resolution multimodal images of a sample. In each published MIST variant \cite{pavlov2020x, alloo2021speckle, pavlov2021directional, alloo2022dark, alloo2023m}, the generality of the algorithm has been expanded by relaxing assumptions about the sample's attenuation, phase, and diffusion characteristics. However, this comes with the trade-off of requiring more input parameters \cite{alloo2023m}, which are necessary to stabilize the algorithm and address particular ill-posed mathematical operations. The most recently published and most general MIST algorithm is presented in Alloo \textit{et al.} \cite{alloo2023m}. Notably, when the required input parameters in this MIST algorithm are chosen correctly, the recovered multimodal images exhibit superior image quality compared to earlier MIST variants \cite{pavlov2020x,alloo2021speckle,alloo2022dark}. Similar to other FPE-derived multimodal retrieval algorithms \cite{pavlov2020x,leatham2023x,ahlers2024x}, the method in Alloo \textit{et al.} requires inverting the Laplacian operator during the multimodal image retrieval process \footnote{In this manuscript, the term `Laplacian operator' refers to the transverse Laplacian operator, which is defined $\nabla_\perp^2=(\partial^2/\partial x^2 + \partial^2/\partial y^2)$. For brevity, the explicit use of `transverse' is omitted throughout.}. The inverse Laplacian operator can be applied in Fourier space using the Fourier derivative theorem \cite{paganin2006coherent}. However, the solution is ill-posed near the Fourier-space origin due to a division-by-zero instability. To mitigate this, a common approach is to apply a Tikhonov regularization \cite{tikhonov1977solutions}, which involves adding a small parameter to the denominator of the operator. This parameter must be carefully selected: if too small, instabilities persist, contaminating the solution with low-frequency artifacts; if too large, low spatial frequency information is lost, and high-spatial frequencies are artificially amplified. Thus, selecting an optimal value for the Tikhonov regularization parameter is crucial for accurately retrieving multimodal signals using algorithms derived from the X-ray FPE. Typically, the value of the Tikhonov regularization parameter in these FPE retrieval algorithms is chosen manually using guesswork, either by inspecting the spatial frequencies near the Fourier-space origin and choosing some suitably small value, or by optimizing image quality metrics of the resultant-regularized image, such as the signal-to-noise ratio. The problem of optimizing the Tikhonov regularization parameter for a given ill-posed problem is 
a topic of broad interest \cite{reichel2013old,park2018parameter, zare2020determination}. For example, one can choose the parameter value corresponding to the point on the L-curve where the curvature is maximized. The L-curve is generated by plotting the seminorm of the regularized solution against the norm of the corresponding residual vector \cite{lawson1995solving,hansen1992analysis,hansen1993use}.

In this work, we present a straightforward yet effective automated iterative algorithm that optimizes the Tikhonov regularization parameter required to stabilize the inverse Laplacian operator. Specifically, we apply this algorithm within our already-developed generalized MIST algorithm (Ref.~\citenum{alloo2023m}), which requires Laplacian inversion to recover a sample's phase image. Our new iterative approach leverages the phase solution from the SBXI TIE-based phase retrieval algorithm developed in Pavlov \textit{et al.} \cite{pavlov2020single}. Notably, this TIE-based algorithm implicitly employs a regularized version of the inverse Laplacian operator, with no need for Tikhonov regularization. As a result, the phase reconstruction is independent of any regularization parameter, and it can serve as a reference for the converged phase in the MIST algorithms.

The iterative method we present aims to minimize the difference between the phase recovered using Pavlov \textit{et al.}'s TIE-based approach and the phase retrieved using the MIST algorithm, which requires Tikhonv regularization \cite{alloo2023m}. Although we develop this method for MIST (i.e., the speckle-based FPE), we anticipate that it can be applied to other experimental techniques, such as propagation-based imaging, where similar FPE models \cite{leatham2023x, leatham2024x,ahlers2024x} and TIE \cite{paganin2002simultaneous} counterparts are also used.
\\
The remainder of this manuscript is organized as follows. The next section will provide a brief overview of the TIE-based phase-retrieval algorithm that was presented in Ref.~\citenum{pavlov2020single} and the FPE-based phase and dark-field retrieval algorithm (MIST) in Ref.~\citenum{alloo2023m}. We will then present the new development in this work: an iterative algorithm designed to optimally regularize the numerical instabilities observed in the MIST algorithm by Alloo \textit{et al.} \cite{alloo2023m}. Following this, we will justify why a multimodal retrieval algorithm (FPE-based) can use an algorithm that only considers X-ray phase shift (TIE-based) to ensure convergence to the optimal and true solution of the inverse problem. We apply the algorithm to experimental SBXI data from a sample composed of four different types of rods, collected using synchrotron light. We present the retrieved phase images with increasing iterations of the new algorithm, along with the final optimized multimodal images (phase and dark-field). We conclude and discuss potential avenues for future research.
 
\section{Theoretical Approach}\label{Theory}
The ill-posed (left) and Tikhonov-regularized (right) forms of the inverse Laplacian operator are,
\begin{equation}
    \nabla_\perp^{-2} = -\mathcal{F}^{-1}\frac{1}{|k_\textbf{r}|^2}\mathcal{F} \rightarrow -\mathcal{F}^{-1}\frac{1}{|k_\textbf{r}|^2+\epsilon}\mathcal{F},
    \label{eqn: InverseLap}
\end{equation}
where the regularized variant has the addition of some small positive regularization parameter $\epsilon$ on its denominator. Above, $\mathcal{F}$ represents the two-dimensional Fourier transform with respect to $x$ and $y$, where $k_x$ and $k_y$ are the associated Fourier-space variables and $|k_\textbf{r}|^2 = k_x^2+k_y^2$.
\\
The sample-induced X-ray phase shift can be computed using the TIE-based approach in Pavlov \textit{et al.} \cite{pavlov2020single}. The approach uses a single set of SBXI data $I_R(\textit{\textbf{r}})$ and $I_S(\textit{\textbf{r}})$, and retrieves the phase $\phi_{TIE}(\textit{\textbf{r}})$ using
\begin{equation}
    \phi_{TIE}(\textit{\textbf{r}}) = \frac{\gamma}{2}\textrm{ln}\left[\mathcal{F}^{-1} \left\{\frac{\mathcal{F}\left\{I_S(\textit{\textbf{r}})/I_R(\textit{\textbf{r}})\right\}}{1+\frac{\gamma \Delta}{2k}|k_\textbf{r}|^2}\right\} \right],
    \label{eqn: PavPhase}
\end{equation}
\begin{figure}[tb]
    \centering
    \includegraphics[width=0.7\linewidth]{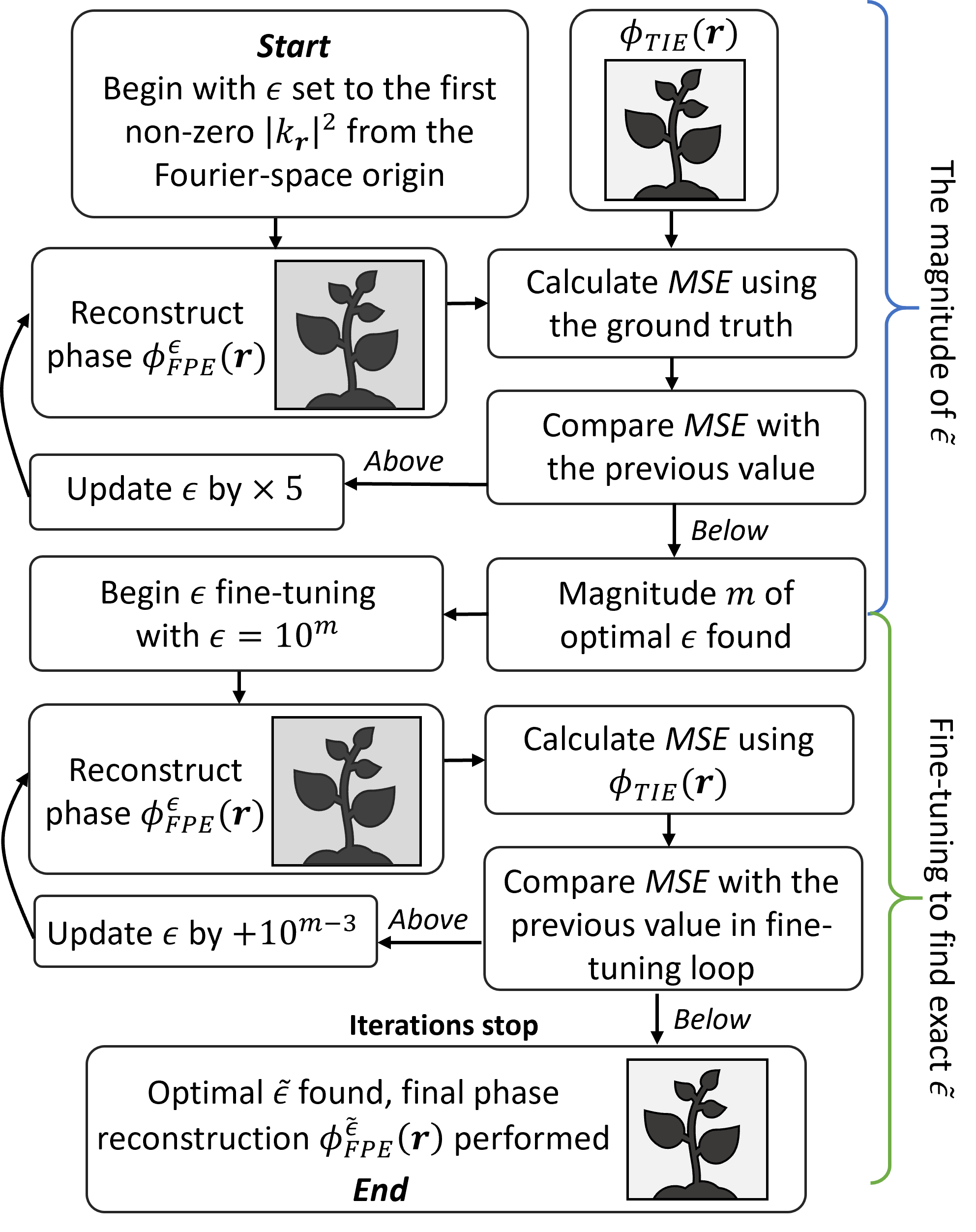}
    \caption{Workflow of the iterative algorithm to optimize the Tikhonov regularization parameter $\epsilon$ for recovering the sample phase using the Multimodal Intrinsic Speckle-Tracking (MIST) in Ref.~\citenum{alloo2023m}. In essence, the algorithm minimizes the mean square error ($MSE$) between the regularization-parameter-independent phase reconstruction $\phi_{TIE}(\textit{\textbf{r}})$ (ground truth) and the MIST phase reconstruction $\phi_{FPE}^{\epsilon}(\textit{\textbf{r}})$ for a given $\epsilon$ value.}
    \label{fig:workflow}
\end{figure}
where $\textbf{r} = (x,y)$ denotes the transverse coordinates perpendicular to the optical axis $z$. In the equation above, $\gamma = \delta / \beta$ (the so-called homogeneous sample assumption), where $\delta$ and $\beta$ are the real and imaginary components of the sample's refractive index, $n = 1 - \delta + i \beta$. The assumption of a constant $\gamma$ value throughout the sample volume implies that the algorithm implicitly couples the sample's attenuation and phase characteristics and, hence, retrieved images. Equation~\ref{eqn: PavPhase} was derived assuming the sample's phase slowly varies with transverse position. The mathematical technique \cite{paganin2002simultaneous} employed in Pavlov \textit{et al.} \cite{pavlov2020single} to recover $\phi_{TIE}(\textit{\textbf{r}})$ gives an implicitly regularized form of the inverse Laplacian operator, i.e., there is a `1 +' term in the denominator of Eq.~\ref{eqn: PavPhase}. 
\\
Equation~\ref{eqn: PavPhase} was derived while neglecting diffuse X-ray scatter from sample structures. To account for these effects and retrieve the complementary dark-field image, the FPE extension of the TIE should be used to model the SBXI forward problem. In this context, a suitable MIST algorithm \cite{pavlov2020x} can be used to solve the inverse problem and recover sample attenuation, phase, and dark-field images. Below, we outline the main steps of the MIST algorithm developed in Ref.~\citenum{alloo2023m}; the reader is referred to Ref.~\citenum{alloo2023m} for a detailed derivation of the algorithm. 
\\
The MIST algorithm assumes the $\gamma$ value is constant throughout the entire sample and that the sample weakly attenuates the X-ray beam. At least four sets of SBXI data---pairs of $I_R(\textit{\textbf{r}})$ and $I_S(\textit{\textbf{r}})$---are required for the algorithm such that a system of four linear equations in terms of $\nabla_\perp^2[(1/k)\phi(\textit{\textbf{r}})-D(\textit{\textbf{r}};\Delta)]$, $D(\textit{\textbf{r}};\Delta)$, $D^x(\textit{\textbf{r}};\Delta)$, and $D^y(\textit{\textbf{r}};\Delta)$ can be generated, as seen in Eq.~6 in Ref.~\citenum{alloo2023m}. The superscripts $x$ and $y$ on the effective diffusion coefficient denote spatial derivatives. Note that pairs of SBXI data can be collected by transversely shifting the membrane during image acquisition and collecting corresponding pairs of reference-speckle and sample-plus-speckle images at each step. The variable $D(\textit{\textbf{r}};\Delta) = 1/2(F(\textit{\textbf{r}})\theta(\textit{\textbf{r}})^2)$ represents the dimensionless FPE effective diffusion coefficient, describing the fraction $F(\textit{\textbf{r}})$ of incident X-rays that are diffusely scattered at an angle $\theta(\textit{\textbf{r}})$. A least-squares solution to the generated system (overdetermined if more than four SBXI data pairs are used) can be computed using Tikhonov-regularized QR decomposition to recover the four unknown variables related to the sample's phase and dark-field. This step also requires selecting an appropriate Tikhonov regularization parameter. Reference~\citenum{alloo2023m} established that a suitable choice for this regularization parameter, based on optimizing image quality, is the standard deviation of the coefficient matrix divided by $10^{4}$. Once the four solutions to the linear system are computed, the sample's \textit{true} effective coefficient $\tilde{D}(\textit{\textbf{r}};\Delta)$ is reconstructed by combining all dark-field-related information from the system's solutions: which is contained in $D(\textit{\textbf{r}};\Delta)$, $D^x(\textit{\textbf{r}};\Delta)$, and $D^y(\textit{\textbf{r}};\Delta)$. This combination is achieved using suitable Fourier-space filtering, as given in Eq.~8 of Ref.~\citenum{alloo2023m}. The Laplacian of the phase can then be retrieved by solving the corresponding SBXI FPE (Eq.~4 in Ref.~\citenum{alloo2023m}) with the retrieved $\tilde{D}(\textit{\textbf{r}};\Delta)$,
\begin{equation}
    \nabla_\perp^{2}\phi_{FPE}(\textit{\textbf{r}}) = \frac{k}{\Delta I_R(\textit{\textbf{r}})}\{I_R(\textit{\textbf{r}}) - I_S(\textit{\textbf{r}}) +\Delta^2\nabla_\perp^{2}[\tilde{D}(\textit{\textbf{r}};\Delta)I_R(\textit{\textbf{r}})]\}.
    \label{eqn: AllooPhase}
\end{equation}
The phase can then be recovered by applying the Tikhonov-regularized inverse Laplacian operator to $\nabla_\perp^{2} \phi_{FPE}(\textit{\textbf{r}})$, i.e., $\nabla_\perp^{-2}\nabla_\perp^{2} \phi_{FPE}(\textit{\textbf{r}})=\phi_{FPE}(\textit{\textbf{r}})$. The inverse Laplacian operator is fundamentally a low-pass Fourier-space filter. The value of $\epsilon$ in its regularized form dictates how much of the low spatial frequencies are altered by the numerical regularization. If chosen too small, the operation is ill-posed and the reconstruction will resemble the numerical instability near the Fourier-space origin (large cloud-like features appear across the entire image). Whereas if it is too large, all of the low spatial frequencies will be suppressed, and the recovered phase will contain edge-enhancement due to propagation-based phase-contrast at sample edges, denoting that the phase-retrieval has not been performed correctly. If this Laplacian inversion is executed accurately, the recovered phase image will contain sample structures and have sharp edges -- not over- or under-smoothed.
\\
The retrieved $\phi_{TIE}(\textit{\textbf{r}})$ from Eq.~\ref{eqn: PavPhase} can be used to optimally select the value of $\epsilon$ required to recover $\phi_{FPE}(\textit{\textbf{r}})$. Recall that the latter phase reconstruction implicitly accounts for diffuse scattering, which has been shown to improve phase recovery quality compared to the TIE-based approach that neglects diffuse scattering \cite{leatham2023x}. An iterative algorithm can be performed to minimize the difference between the retrieved $\phi_{FPE}(\textit{\textbf{r}})$ for a given $\epsilon$ (denoted by $\phi_{FPE}^\epsilon(\textit{\textbf{r}})$ in the equations below) and $\phi_{TIE}(\textit{\textbf{r}})$, as measured by the mean square error $MSE$, 
\begin{equation}
    MSE = \frac{1}{p}\sum_{i=1}^{p}(\phi_{TIE:\:i}-\phi_{FPE:\:i}^\epsilon)^2,
    \label{eqn: MSE}
\end{equation}
where $i$ denotes the $i$-th pixel position in an image consisting of a total of $p$ pixels. The MSE provides a metric for how close, globally, the recovered $\phi_{FPE}^\epsilon(\textit{\textbf{r}})$ values are to the ground truth $\phi_{TIE}(\textit{\textbf{r}})$ values. The iterative algorithm starts with an initial $\epsilon$ equal to the first non-zero value in $|k_\textbf{r}|^2$ from the Fourier-space origin, as we expect $\epsilon$ to be relatively small compared to the spatial frequencies within the vicinity of the origin. Next, the algorithm determines the approximate magnitude of the optimal $\epsilon$ value by minimizing the MSE between the recovered $\phi_{FPE}^\epsilon(\textit{\textbf{r}})$ and $\phi_{TIE}(\textit{\textbf{r}})$. If the MSE increases compared to the previous iteration, the previous $\epsilon$ magnitude is considered optimal; otherwise, $\epsilon$ is updated to a value five times larger. The top portion of Fig.~\ref{fig:workflow} marked in blue shows the iterative algorithm's workflow when determining the optimal magnitude of $\epsilon$. Once determining the approximate magnitude $m$, fine-tuning of $\epsilon$ occurs, and this is summarised in the bottom portion of Fig.~\ref{fig:workflow} marked in green. To begin the fine-tuning, the algorithm calculates the MSE between $\phi_{TIE}(\textit{\textbf{r}})$ and $\phi_{FPE}^\epsilon(\textit{\textbf{r}})$ reconstructed with an $\epsilon$ value of $10^{m}$. Subsequently, $\epsilon$ is increased in increments of $10^{m-3}$, and the $\phi_{FPE}^\epsilon(\textit{\textbf{r}})$ and corresponding MSE are calculated at each step. This fine-tuning process ceases when the current iteration's MSE exceeds that of the previous iteration. The value of $\epsilon$ at the previous iteration is taken as the optimal regularization parameter value $\tilde{\epsilon}$, and this retrieves the optimal phase image $\phi_{FPE}^{\tilde{\epsilon}}(\textit{\textbf{r}})$. A fine-tuning step size in $\epsilon$ of $10^{m-3}$ was chosen because the percentage difference between two $\phi_{FPE}^\epsilon(\textit{\textbf{r}})$ reconstructions differing by this amount was only 0.14\%, demonstrating that this step-size effectively ensures the optimal parameter value is found. 

Up until this point, the MIST algorithm has neglected sample X-ray attenuation. To extend to weakly attenuating samples, the sample's attenuation term $t_0^{\tilde{\epsilon}}(\textit{\textbf{r}})$ is calculated using a result of the projection approximation $t_0^{\tilde{\epsilon}}(\textit{\textbf{r}}) = \textrm{exp}[2/\gamma\:\phi_{FPE}^{\tilde{\epsilon}}(\textit{\textbf{r}})]$. The recovered dark-field $\tilde{D}(\textit{\textbf{r}};\Delta)$ from the solved system of linear equations (which assumed no attenuation) can then be attenuation-corrected by dividing by $t_0^{\tilde{\epsilon}}(\textit{\textbf{r}})$ to give the attenuating-object approximation $\tilde{D}_A^{\tilde{\epsilon}}(\textit{\textbf{r}};\Delta)$.\\
We want to conclude this section by justifying the approach taken here of using an algorithm that only considers X-ray attenuation and phase to optimize one that also accounts for diffuse X-ray scatter. Put differently, why we use $\phi_{TIE}(\textit{\textbf{r}})$ to optimize $\phi_{FPE}^{\epsilon}(\textit{\textbf{r}})$? This can be clarified by examining the difference between the TIE and FPE models, as well as the Fourier-space spatial frequencies most sensitive to changes in the Tikhonov regularization parameter. The FPE extends the TIE by incorporating a diffusion term that accounts for sample-imposed diffuse X-ray scattering--specifically, sample-imposed phase variations that are smaller than the imaging system's spatial resolution. Near the Fourier-space origin ($|k_\textbf{r}|^2 = \textbf{0}$), the distinction between the TIE and FPE models diminishes as information about diffuse X-ray scatter is not present at low-Fourier-space spatial frequencies. Both the FPE and TIE models contain primarily attenuation information at these low spatial frequencies. Angles associated with phase and dark-field effects are small, and hence, the changes in contrast are very local in the image plane; thus, information regarding these effects is contained at high spatial frequencies in Fourier space. This is explicitly seen and exploited in the Fourier-based single-grid phase and dark-field retrieval approach in Wen \textit{et al.} \cite{Wen2010}. The regularization parameter $\epsilon$ in Eq.~\ref{eqn: InverseLap} adjusts the Fourier frequencies near the origin, where numerical instabilities in the inverse Laplacian operator can occur. Therefore, to optimize the value of $\epsilon$ during the phase retrieval step of the MIST approach in Alloo \textit{et al.} \cite{alloo2023m}, we only need an image of the sample's low spatial frequency features (attenuation or phase image) as the ground truth. While the attenuation image recovered using Beer's law \cite{paganin2006coherent} could also serve as the ground truth in this algorithm, a TIE-retrieved result is preferable as these methods have been shown to suppress noise and remove propagation-based edge fringes, thereby improving the phase-retrieved image's quality \cite{gureyev2017unreasonable}. Additionally, the phase retrieval method in Pavlov \textit{et al.} \cite{pavlov2020single} is suitable as it has demonstrated quantitative accuracy in its retrieved images \cite{rouge2021comparison,rosich2024exploring}.
\begin{figure}[tbh]
    \centering
    \includegraphics[width=0.677\linewidth]{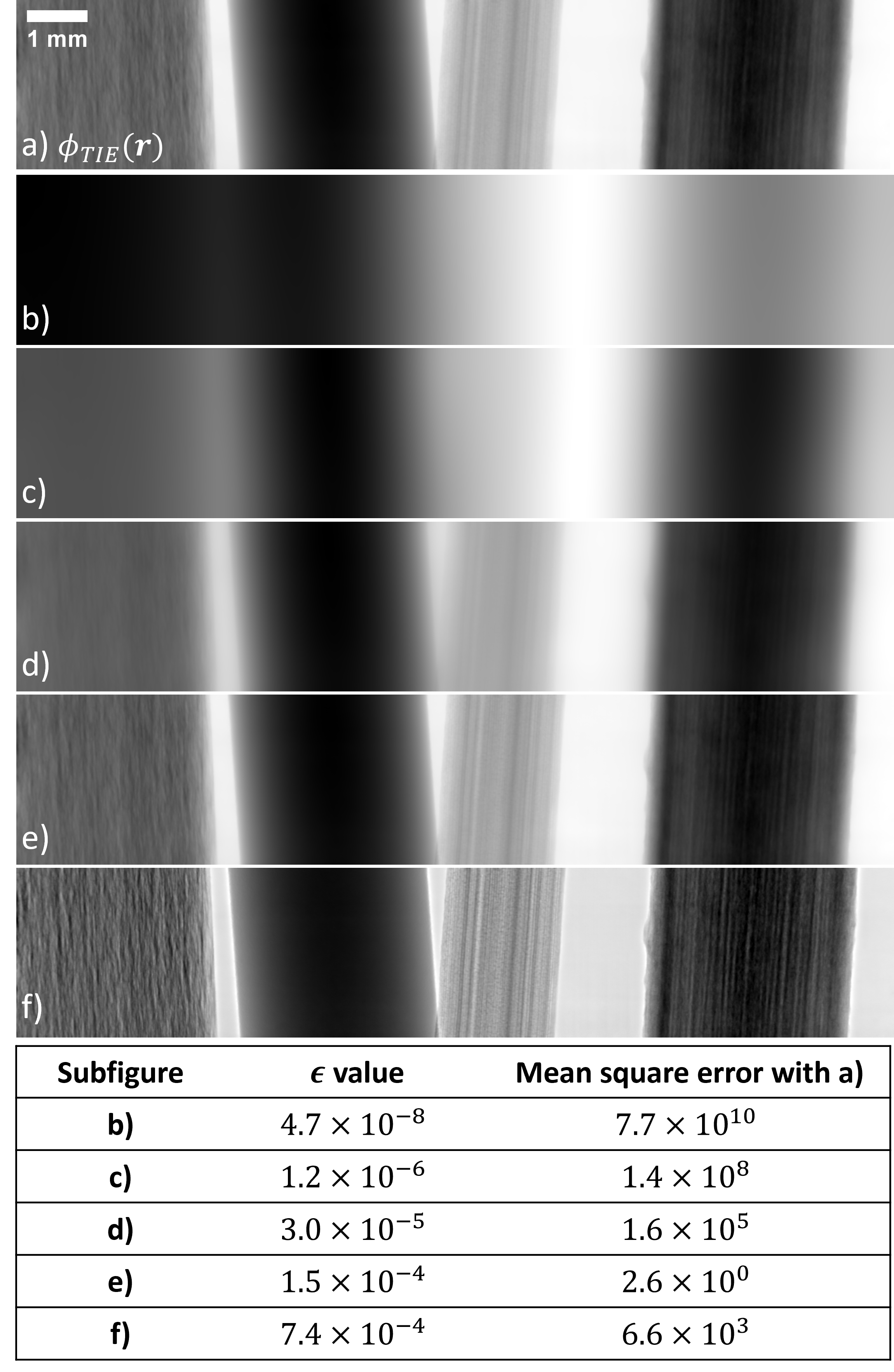}
    \caption{Retrieved phase shifts for the four-rod sample: a) using the transport-of-intensity equation (TIE)-based algorithm in Ref.~\citenum{pavlov2020single} (ground truth, stable retrieval, no regularization required), and b)--f) using the MIST algorithm in Ref.~\citenum{alloo2023m} with varying Tikhonov regularization parameters $\epsilon$. The MSE of b)--f) compared to $\phi_{TIE}(\textit{\textbf{r}})$ is provided in the table at the bottom of the figure. Images are displayed on a linear grayscale range with [min, max] rads: a) [-163.9, -1.9], b) [-2.9, -2.5]$\times 10^{5}$, c) [-45.9, -6.4]$\times 10^{2}$, d) [-497.0, -9.7], e) [-51.4, 2.1], and f) [-6.5, 1.9]. }
    \label{fig: PhaseDifferentEpsilon}
\end{figure}
\section{Experimental Methods}
The presented iterative algorithm was tested on an SBXI data set of a four-rod sample imaged in the first experimental hutch, located 24.0 m from the bending magnet source, of the Australian Synchrotron's MicroCT beamline. This data was first published in Alloo \textit{et al.} \cite{alloo2024} and is available within. The four-rod sample is comprised of four materials (from left to right in all of the images shown later in the manuscript): a reed diffuser stick, a polymethyl methacrylate (PMMA) rod, a toothpick, and a tree twig. The X-ray beam used for the SBXI experiments was a 25 keV monochromatic beam, generated using a double-multilayer monochromator, with a spectral bandwidth of $\Delta E/E \approx 3\times10^{-2}$. The speckle membrane consisted of two layers of P800 grit sandpaper, which was positioned 0.3 m upstream of the four-rod sample. The detector system was a pco.edge 5.5 complementary metal-oxide-semiconductor (CMOS) camera with 2560$\times$2160 pixels, each pixel measuring 6.5 $\SI{}{\micro\meter}$. This camera was coupled to a GGG:Eu/Tb scintillator with a 1 $\times$ optical lens placed in between. This detector system was located $\Delta = 0.7$ m downstream of the sample, where the sandpaper-generated speckle field had an effective speckle size of 18.3 $\SI{}{\micro\meter}$ and a Michelson visibility of 0.28. In total, 13 pairs of SBXI images were acquired by translating the sandpaper perpendicular to the optical axis, capturing reference-speckle and sample-plus-speckle images at each step.
\begin{figure}[tbh]
    \centering
    \includegraphics[width=0.7\linewidth, trim=5 5 5 5,clip]{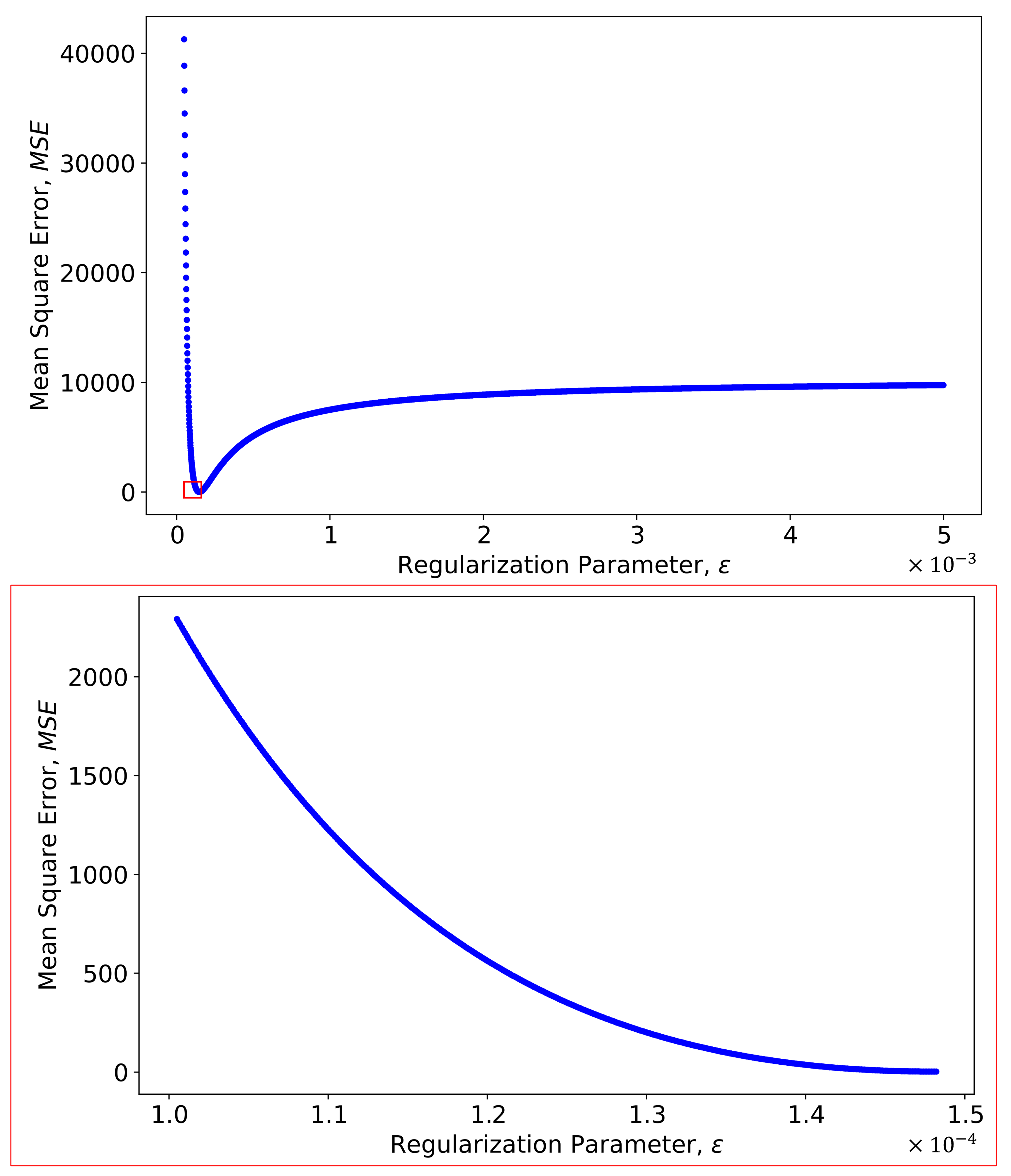}
    \caption{(top) MSE between retrieved phase $\phi_{FPE}^{\epsilon}(\textit{\textbf{r}})$ for a given Tikhonov regularization parameter $\epsilon$ (as defined in Eq.~\ref{eqn: InverseLap}) and the ground $\phi_{TIE}(\textit{\textbf{r}})$. The plot was generated by reconstructing $\phi_{FPE}^{\epsilon}(\textit{\textbf{r}})$ starting at $\epsilon = 4.9\times 10^{-5}$, with increments of $\times 10^{-6}$, and ending at $\epsilon = 4.9\times 10^{-3}$, calculating the MSE at each step. (bottom) A portion of the plot above (indicated in red) showing the range of $\epsilon$ values evaluated by the iterative algorithm within the fine-tuning step of the workflow illustrated in Fig.~\ref{fig:workflow}.}
    \label{fig: MSEPlots}
\end{figure}
\section{Results and Discussion}
\begin{figure}[tb]
    \centering
    \includegraphics[width=0.9\linewidth]{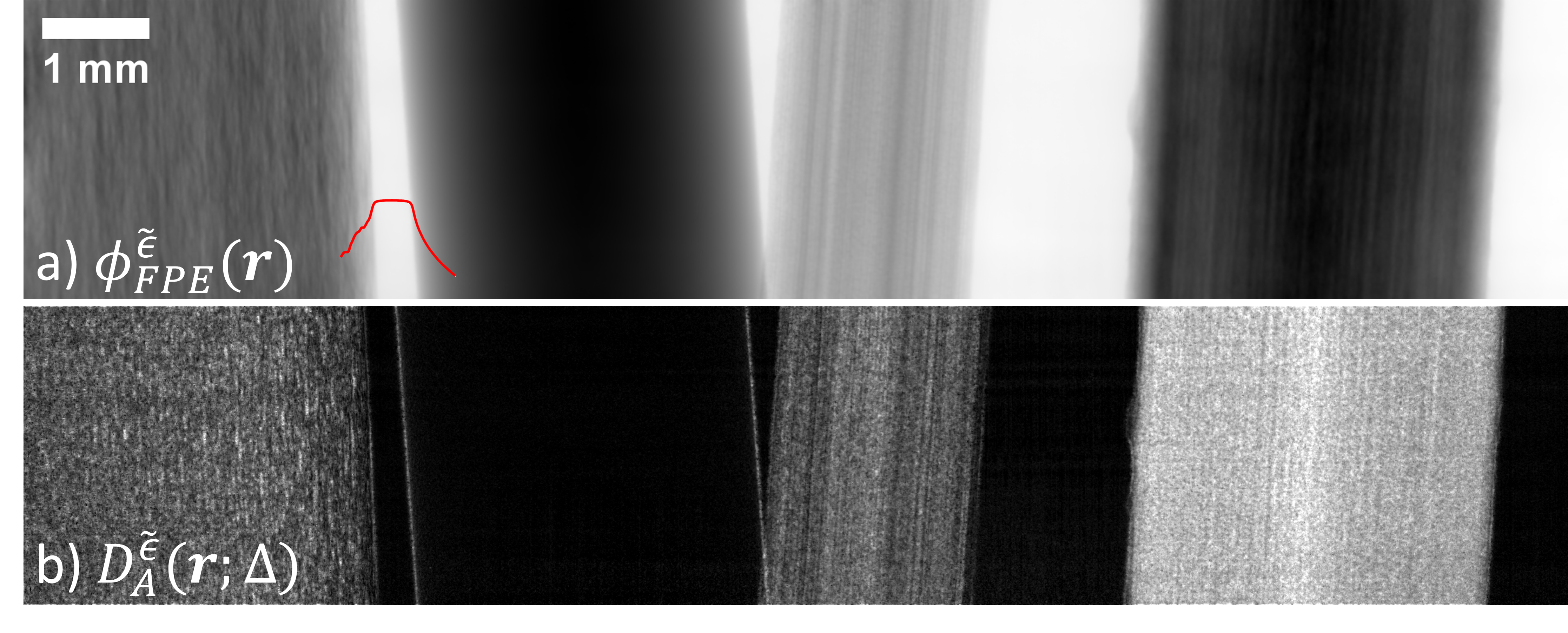}
    \caption{Optimized a) phase and b) dark-field reconstructions of the four-rod sample using a Tikhonov regularization parameter of $\tilde{\epsilon}= 0.0001481$, as determined by our presented iterative algorithm. The red line in a) shows the recovered $\phi_{FPE}^{\tilde{\epsilon}}(\textit{\textbf{r}})$ values at the specified location, spanning the reed diffuser stick into air and then into the polymethyl methacrylate (PMMA) rod. The plot demonstrates optimal regularization, with no under- or over-smoothing at interfaces. Images are displayed on a linear grayscale range with [min, max]: a) [-160.3, -2.0] rads and b) [0.0, 4.4]$\times 10^{-11}$.}
    \label{fig: OptimalRecons}
\end{figure}

Using the iterative MIST algorithm described in Sec.~\ref{Theory}, the retrieved phase and, subsequently, dark-field images of the four-rod sample were optimized. Figure~\ref{fig: PhaseDifferentEpsilon} shows the `ground truth' phase image (Fig.~\ref{fig: PhaseDifferentEpsilon}a) retrieved using the TIE-based phase-retrieval approach \cite{pavlov2020single}, alongside the MIST-retrieved phase images generated with different regularization parameters during the Laplacian inversion step (Figs.~\ref{fig: PhaseDifferentEpsilon}b--f). 
Specifically, Figs.~\ref{fig: PhaseDifferentEpsilon}b--f show the phase reconstructions obtained during the iterations of the algorithm focused on determining the magnitude of the optimal $\epsilon$ value. When $\epsilon$ is too small, numerical instabilities near the Fourier-space origin amplify certain low spatial frequencies, causing the phase image to appear cloudy, as seen in Fig.~\ref{fig: PhaseDifferentEpsilon}b. This cloudiness diminishes with increasing $\epsilon$, as demonstrated in Figs.~\ref{fig: PhaseDifferentEpsilon}c--f. However, Fig.~\ref{fig: PhaseDifferentEpsilon}f illustrates that when $\epsilon$ becomes too large, low spatial frequencies are overly suppressed, leaving strong phase contrast (propagation-based Fresnel fringes) at interfaces within the sample. The optimal magnitude of $\epsilon$, the first step in the iterative algorithm, was found to be on the order $\times 10^{-4}$, as shown in Fig.~\ref{fig: PhaseDifferentEpsilon}e and the table at the bottom of Fig.~\ref{fig: PhaseDifferentEpsilon}. Fine-tuning of $\epsilon$ was performed around the determined optimal magnitude of $\times 10^{-4}$. The upper plot in Fig.~\ref{fig: MSEPlots} illustrates how the MSE between $\phi_{FPE}^{\epsilon}(\textit{\textbf{r}})$ and $\phi_{TIE}(\textit{\textbf{r}})$ varies with the regularization parameter. Notably, this plot was generated over a much wider range of $\epsilon$ values than those trialed during the iterative algorithm's fine-tuning step. This plot confirms that minimizing the MSE between $\phi_{FPE}^{\epsilon}(\textit{\textbf{r}})$ and $\phi_{TIE}(\textit{\textbf{r}})$ is an effective metric for determining the optimal $\epsilon$ value, i.e., there is a single global minimum. The bottom plot in Fig.~\ref{fig: MSEPlots} shows the actual range of $\epsilon$ values trialed during the fine-tuning step, with the range of this plot denoted by the red box in the plot above. This plot shows that the algorithm ceases once the MSE has been minimized (the plot ends at the minimum), i.e., the MSE of the next fine-tuning iteration is larger than the current, so it is the current one that is the global MSE minimum. This global MSE minimum for the four-rod sample was located at 
$\tilde{\epsilon}= 0.0001481$, yielding the optimal Tikhonov regularization parameter, and hence, phase image reconstruction $\phi_{FPE}^{\tilde{\epsilon}}(\textit{\textbf{r}})$. The entire Python script implementing the TIE phase retrieval \cite{pavlov2020single} and the iterative Tikhonov regularization optimization variant of the MIST algorithm \cite{alloo2023m} (489 iterations in total) completed in 140 seconds on a 13th Gen Intel(R) Core(TM) i7-13800H 2.5 GHz computer with 32 GB of RAM. This script and the SBXI data of the four-rod can be found in an open GitHub repository \cite{reconstruction_Github}. 
\\
The optimized phase image and the corresponding dark-field image extracted using the MIST algorithm are shown in Figs.~\ref{fig: OptimalRecons}a) and b), respectively. It is important to reiterate here that the TIE algorithm (Pavlov \textit{et al.} \cite{pavlov2020single}) neglects X-ray diffusion within the sample. Only MIST algorithms, one of which is optimized in this work, can recover this information through the dark-field image. The red trace in Fig.~\ref{fig: OptimalRecons}a shows the recovered $\phi_{FPE}^{\tilde{\epsilon}}(\textit{\textbf{r}})$ values at the specified location. This line profile traces from the reed diffuser stick into air and then into the PMMA rod. There is no residual phase contrast at the interfaces between neighboring materials in the four-rod's retrieved phase image, and the edges are recovered sharply without over-smoothing.  This plot demonstrates that the Tikhonov regularization parameter $\epsilon$ in the inverse Laplacian operator has indeed been optimized, allowing for an accurate reconstruction of the four-rod’s phase. The retrieval algorithm must address all contrast channels in tandem to accurately retrieve information from a sample with attenuating, refracting, and diffusing characteristics. For the X-ray FPE, precise phase retrieval is essential to ensure the accurate reconstruction of the dark-field signal, and vice versa, as the two channels are inherently coupled. As demonstrated, the presented iterative algorithm refines the phase retrieval step, ensuring accurate phase and dark-field recovery.

\section{Conclusion}
In this work, we developed an automated iterative algorithm to optimize the Laplacian inversion step required in FPE-based multimodal image retrieval algorithms. Specifically, we applied this to phase recovery within the most generalized variant of our developed Multimodal Intrinsic Speckle-Tracking (MIST) algorithm \cite{alloo2023m}. The inverse Laplacian operator required in MIST is evaluated using the Fourier-derivative theorem, but this operation becomes unstable near the Fourier-space origin. If not rectified numerically--such as by applying a suitable Tikhonov regularization--these instabilities will contaminate the recovered phase image with low spatial frequency artifacts. Our iterative algorithm optimizes the required Tikhonov regularization by utilizing the phase image retrieved via a transport-of-intensity equation (TIE)-based approach \cite{pavlov2020single}, which employs an inherently stable inverse Laplacian operator that requires no regularization, meaning there are no input parameters to fine-tune. We successfully applied the algorithm to experimental speckle-based X-ray imaging (SBXI) data of a sample comprised of a reed diffuser stick, a PMMA rod, a toothpick, and a tree twig. This optimization yielded phase and dark-field images of the four rods, revealing complementary information not attainable with the TIE-based approach alone. We trust that the developed regularization-optimization algorithm makes the MIST retrieval algorithms more convenient for users of SBXI, as it eliminates the need for manual parameter selection. Future research could enhance the rate of convergence of the iterative algorithm by incorporating more advanced fine-tuning techniques than those presented here. We anticipate that this work will be beneficial for other FPE-based retrieval algorithms, such as those employed in free-space propagation techniques \cite{ahlers2024x,leatham2023x}, as they also require the inversion of the Laplacian operator to recover phase and dark-field images of a sample.

\section*{Acknowledgements}
The authors acknowledge valuable discussions with David M. Paganin. We also thank Michelle K. Croughan and Jannis N. Ahlers for collecting the speckle-based X-ray imaging data of the four-rod sample at the MicroCT Beamline at the Australian Synchrotron (proposal 19663). Additionally, we acknowledge funding support from the Australian Research Council (FT18010037 and DP230101327).

\bibliographystyle{unsrt}  
\bibliography{refs}

\end{document}